\documentclass[review]{elsarticle}

\usepackage{lineno,hyperref}
\usepackage{amsmath}
\usepackage{amsfonts}
\usepackage{amssymb}
\usepackage{subfigure}
\usepackage[margin=0.75in]{geometry}
\usepackage{multirow}

\journal{Journal of Magnetism and Magnetic Materials}

\begin{document}
\begin{frontmatter}

\title{Skyrmion Based Spin-Torque Nano-Oscillator}

\author{Debasis~Das, Bhaskaran~Muralidharan and~Ashwin~Tulapurkar}
\address{Department of Electrical Engineering, IIT Bombay,
    Mumbai 400076, India}

%
\cortext[mycorrespondingauthor]{Corresponding author\\ Debasis Das (ddas@ee.iitb.ac.in)}

%

\begin{abstract}
Using micromagnetic simulation, we investigate the self-sustained oscillation of magnetic skyrmion in a ferromagnetic circular nanodot, driven by spin-torque which is generated from a reference layer of a circular nanopillar device. We demonstrate, by lowering the value of uniaxial anisotropy constant ($K_u$), the velocity of the skyrmion can be increased and using this property, gyration frequency of the skyrmion oscillator can be enhanced. Annihilation of the skyrmion at higher current densities, limit the gyration frequency of the oscillator, whereas by modifying the $K_u$ value at the edge of nanodot, we are able to protect the skyrmion from being annihilated at higher current densities which in turn, increases the gyration frequency of the skyrmion based oscillator. By linear fitting the velocity value, obtained from the motion of the skyrmion in a nanostrip, we also predict the gyration frequency of the skyrmion in the nanodot which proves the validity of our idea in an intuitive way. We have also varied the radius of the nanodisk to see its effect on skyrmion.
\end{abstract}

\begin{keyword}
Skyrmion, spin-transfer torque, oscillator.
\end{keyword}

\end{frontmatter}


\section{Introduction}
Skyrmion is a soliton solution in the area of non-linear field theory proposed by Tony Skyrme \cite{skyrme1962unified}, which has a stable topology. Skyrmion in a magnetic system represents a stable vortex-like spin texture which has an integer topological number \cite{nagaosa2013topological}. This topological number or skyrmion number is given by 
\begin{equation}
Q=\frac{1}{4\pi}\int \int \boldsymbol{m} \cdot \left( \frac{\partial \boldsymbol{m}}{\partial x} \times \frac{\partial \boldsymbol{m}}{\partial y} \right) dxdy
\end{equation} 
where, $\boldsymbol{m}$ is normalized magnetization vector. Due to this topological protection \cite{nagaosa2013topological,sampaio2013nucleation,zhou2014reversible}, it is very difficult to destroy the skyrmion which shows a possibility to use it as an information bit. Skyrmion were originally discovered in bulk ferromagnet (FM) \cite{pfleiderer2010skyrmion} such as MnSi \cite{jonietz2010spin,muhlbauer2009skyrmion}, $\mathrm{Fe_{0.5}Co_{0.5}Si}$ \cite{yu2010real} where inversion symmetry is broken, and later it was created in thin FM layer with a perpendicular magnetic anisotropy, grown on heavy metal \cite{jiang2015blowing,woo2016observation,yu2016room,romming2013writing,leonov2016properties,romming2015field}. In recent years, skyrmion gained a lot of attention due to various reasons such as (i) small size (diameter with few nanometer) \cite{rohart2013skyrmion,moreau2016additive}, which is useful for higher density storage, (ii) topological stability, which prevents the skyrmion (information) from being lost and (iii) lower depinning current density($\sim10^6$ A/$\mathrm{m^2}$ in B20 structure) \cite{jonietz2010spin,yu2012skyrmion} than that of domain wall ($\sim 10^{12}$ A/$\mathrm{m^2}$)\cite{yang2015domain}. Dzyaloshinskii-Moriya interaction (DMI) plays an important role to stabilize the skyrmion in the FM layer \cite{rohart2013skyrmion,fert2013skyrmions}. In magnetic materials, Heisenberg exchange interaction favors the collinear alignments of spins whereas DMI favors the non-collinear alignment. Competition between these two interactions leads to the formation of skyrmion in FM \cite{fert2013skyrmions}. Due to these above-mentioned advantages, devices such as skyrmion-based racetrack memory \cite{kang2016complementary,tomasello2014strategy,zhu2018skyrmion}, logic gates \cite{zhang2015magnetic,luo2018reconfigurable}, synaptic devices \cite{huang2017magnetic,li2017magnetic} were proposed using skyrmion motion. Although theoretical calculations were done at a low temperature, some experimental works have shown the stability of the skyrmion at room temperature \cite{yu2016room,maccariello2018electrical,lin2018observation}.  Annihilation of the skyrmion due to Magnus force \cite{sampaio2013nucleation,kang2016voltage} remained as an obstacle for designing such skyrmion based devices. Due to the skyrmionic Hall effect \cite{nagaosa2013topological}, skyrmion follows a curved path during its motion, until it reaches near the edge which helps to design the skyrmion-based spin torque oscillators \cite{zhang2015current,garcia2016skyrmion}. In this type of oscillator, skyrmion moves in a circular path whose diameter is determined by the combined effect of the spin-transfer torque, Magnus force, and the edge repulsion force acting on it. As skyrmion oscillation frequency depends on the skyrmion velocity which is directly proportional to current density \cite{sampaio2013nucleation}, so to achieve high-frequency one needs to apply higher current density. This higher velocity increases the Magnus force on the skyrmion, which pushes it further towards the boundary that finally annihilates the skyrmion. Thus, the frequency of the oscillator gets limited due to the skyrmionic Hall effect, which can only be overcome, if we are able to protect the skyrmion from being annihilated at higher current densities.

In this article, we explore the self-sustained oscillation of skyrmion results from the vortex-like spin current in a circular nanopillar geometry using micromagnetic simulation. In section \ref{theory}, we describe the device structure, mathematical formulation, and simulation details. Following this, we explore how uniaxial anisotropy constant, affects the motion of skyrmion in a nanostrip, in section \ref{Ku_dendency}. In section \ref{osci_section}, we discuss in details about the skyrmion-based oscillator and we also demonstrate how to protect the skyrmion in the nanodot from being annihilated so that the frequency of the oscillator can be increased further. In section \ref{ku_effect}, we explore the effect of uniaxial anisotropy constant on the oscillation frequency ($f$) and the radius ($r_{Sk}$) of the skyrmion. We also discuss a model which uses a linear fitting to predict the frequency of the oscillator using the velocity value obtained from the skyrmion motion in a nanostrip. Finally, in section \ref{nanodisk_diameter}, we show how the radius of the nanodisk affects the gyration frequency and the radius of the skyrmion.

\section{Theoretical Formulation}\label{theory}
\subsection{Device Structure}\label{device_structure}

\begin{figure}[h]
    \centering
    \subfigure[]{\includegraphics[scale=0.25]{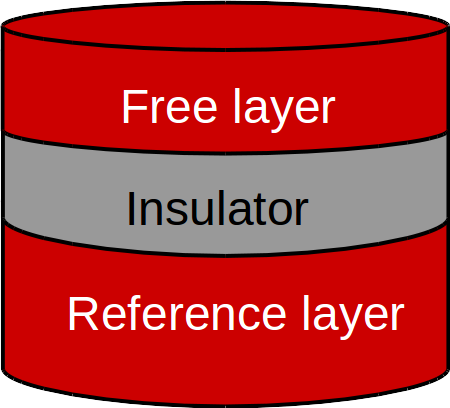}}
    \hspace*{1.5cm}
    \subfigure[]{\includegraphics[scale=0.23]{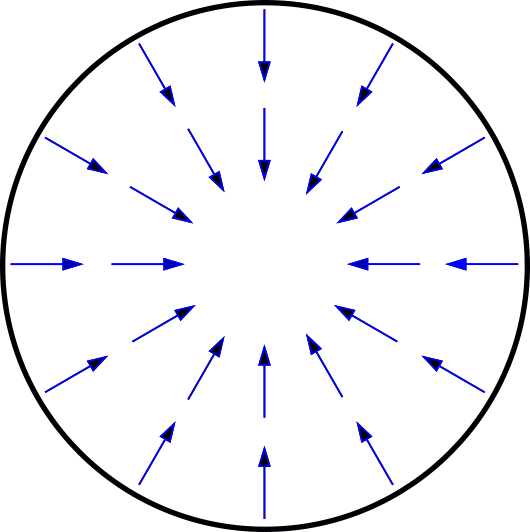}}
    
    \caption{(a) Schematic of the skyrmion based oscillator device. RL and the FL are separated by an insulator layer. (b) Vortex-like magnetization pattern of the RL which creates the spin current with vortex-like polarization. The spin current gyrates the skyrmion in the FL.}
    \label{device}
\end{figure}
In this section, we describe the device structure as shown in Fig. \ref{device} (a), where two FM layers are separated by an insulator layer. The top FM layer, called the free layer (FL) has uniaxial perpendicular magnetic anisotropy with DMI such that skyrmion can be stabilized in this layer. The bottom layer is the reference layer (RL) where the magnetization is fixed and it is assumed to have an in-plane vortex-like spin polarization as shown in Fig. \ref{device} (b). The electrons are assumed to travel from RL to FL where the spin-polarized current takes the form of the vortex and due to spin-transfer torque, skyrmion gyrates in the FL. The spin polarization of the RL is assumed to be $\textbf{m}_p=\left( cos\phi,sin\phi,0\right)$, where $\phi=tan^{-1}(\frac{y}{x})+\psi$. Here, ($x$,$y$) is the spatial coordinate with the origin being at the center of the nanodot and the $\psi$ is assumed to $\pi$ in our simulation. Although the device structure is similar to the device, shown in Ref. \cite{garcia2016skyrmion}, but we modified the device (shown in Fig. \ref{circular_barrier}) to enhance the gyration frequency of the skyrmion in the FL. We create an annular region of high anisotropy around the perimeter of the nanodot, to prevent the skyrmion annihilation at higher current densities.

\subsection{Mathematical Model}
The proposed device is simulated by solving the Landau-Lifshitz-Gilbert-Slonczewski (LLGS) equation described as follows
\begin{equation}\label{llgs}
\frac{d\bold{m}}{dt}=-|\gamma|\bold{m}\times\bold{H}_{eff}+
\alpha \left( \bold{m}\times \frac{d\bold{m}}{dt}\right) +
|\gamma|\beta \left( \bold{m}\times \bold{m}_p\times \bold{m}\right) 
\end{equation}
where, $\bold{m}=\bold{M}/M_S$ is the normalized magnetization vector and $M_S$ is the saturation magnetization. $|\gamma|$ is the gyromagnetic ratio and $H_{eff}$ is the effective magnetic field given by 
\begin{equation}\label{Hfield}
H_{eff}=-\frac{1}{\mu_0 M_s} \frac{\delta E}{\delta \bold{m}}
\end{equation}
where, $\mu_0$ is the free space permeability and $E$ is the free energy and is given by \cite{zhou2014reversible,rohart2013skyrmion}
\begin{equation}\label{energy}
E = \int dV\left[ A\left( \nabla \bold{m}\right) ^2+K_u\left[ 1-\left( \bold{m}\cdot \bold{z}\right)^2\right]-\frac{\mu_0}{2}\bold{m}\cdot \bold{H}_d+\varepsilon_{DMI} \right]
\end{equation}
Here, the first term denotes the exchange energy and $A$ is the exchange stiffness constant. The second term denotes the uniaxial anisotropy energy and $K_u$ is the uniaxial anisotropy constant. The third term is due to the demagnetization field $\bold{H}_d$ and the final term denotes the interfacial form of DMI, which leads to N\'{e}el-type skyrmion, is given by 
\begin{equation}\label{dmi}
\varepsilon_{DMI}=D\left( m_z\frac{\partial m_x}{\partial x}+m_z\frac{\partial m_y}{\partial y}-m_x\frac{\partial m_z}{\partial x}-m_y\frac{\partial m_z}{\partial y}\right)
\end{equation}
where, $D$ is the DMI constant. The third term of Eq. \ref{llgs} represents the Slonczewski spin transfer torque, where the factor $\beta$ is given by $\beta=\hbar P J/(2\mu_0 e t M_s)$. Here $\hbar$, $P$ and $J$ are reduced Planck's constant, spin polarization and the current density respectively. $e$ is the electronic charge  and $t$ represents thickness of the FL.

The motion of the skyrmion in FL is described by Thiele \cite{thiele1973steady} equation, given by \cite{garcia2016skyrmion}
\begin{equation}\label{thiele}
\bold{G}\times \bold{v}+\alpha D_0 \bold{v}+\Gamma_{STT}=-\frac{\partial E}{\partial \bold{X}}
\end{equation}
which describes different forces acting on the skyrmion. The first term of Eq. \ref{thiele} reperesents Magnus force, where, \textbf{G} is gyrovector and \textbf{v} is the velocity of the skyrmion. The gyrovector is defined by 
\begin{equation}
G=\frac{M_s}{\gamma}\int dV sin(\theta)\left( \nabla \theta \times \nabla \phi\right)
\end{equation}
where, the $\theta$ and $\phi$ are the polar and azimuthal angle. It can be shown that \cite{kang2016skyrmion}, the gyrovector can be written as  $\boldsymbol{G}=4\pi Q \hat{z}$. The second term represents the damping term, where $D_0$ is the diagonal components of the damping tensor $D_{ij}$, given by
\begin{equation}
D_{ij}=\frac{M_s}{\gamma}\int dV\left( \frac{\partial \theta}{\partial x_i} \frac{\partial \theta}{\partial x_j}+sin^2(\theta) \frac{\partial \phi}{\partial x_i} \frac{\partial \phi}{\partial x_j} \right)
\end{equation}
where, $x_i$ represents the cartesian coordinates $\left( x,y,z\right) $. The third term $\Gamma_{STT}$ of Eq. \ref{thiele} represents the spin torque term which can be derived by Rayleigh dissipation function \cite{garcia2016skyrmion,kim2012spin}. The term on the right hand side of Eq. \ref{thiele} represents the force acting on the skyrmion.

\subsection{Simulation Details}\label{simdetails}
The proposed device is simulated using Object Oriented MicroMagnetic Framework (OOMMF) public code \cite{oommf} by incorporating the DMI extension module \cite{dmi}. The radius of the FL is assumed to be 50 nm with a thickness of 0.4 nm which is discretized into the cell size of $1\times1\times0.4~\mathrm{nm^3}$. The material parameters were adopted from Ref. \cite{sampaio2013nucleation} which includes gyromagnetic ratio $\gamma$=2.211$\times~10^5$ m/(A.s), Gilbert damping coefficient $\alpha$=0.3, exchange stiffness $A$=15 pJ/m, saturation magnetization $M_S$=580$\times10^3$ A/m, spin polarization P=0.4 and the interfacial DMI constant $D$=3 mJ/$\mathrm{m^2}$. In this simulation, we have neglected the dipolar coupling due to ultrathin FM layer\cite{rohart2013skyrmion}. We have chosen two different values of perpendicular magnetic anisotropy $K_u$=0.6 and 0.8 MJ/$\mathrm{m^3}$ to investigate the effect of the anisotropy constant on the gyration frequency. 

\section{Result \& Discussion}
\subsection{$K_u$ Dependency on Skyrmion} \label{Ku_dendency}
\begin{figure}[h]
    
    \subfigure[]{\includegraphics[scale=0.14]{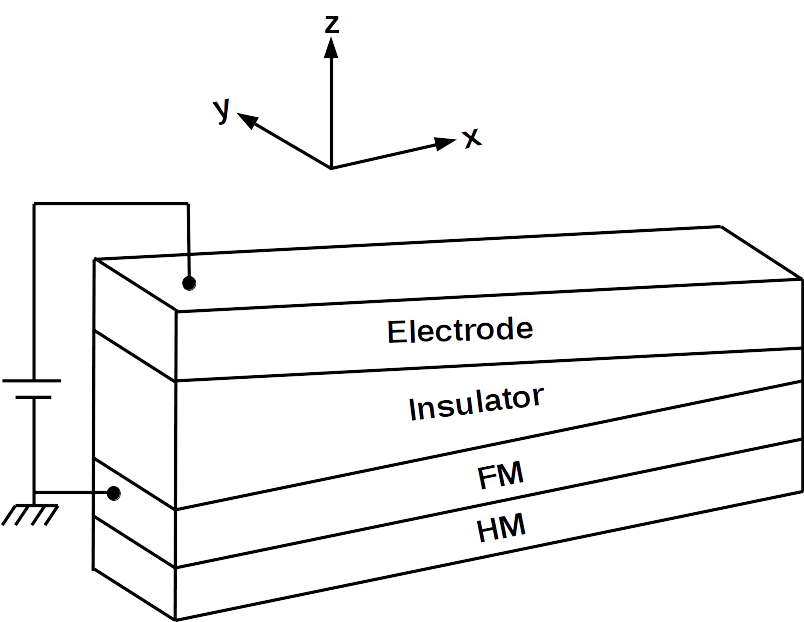}}
    \subfigure[]{\includegraphics[scale=0.17]{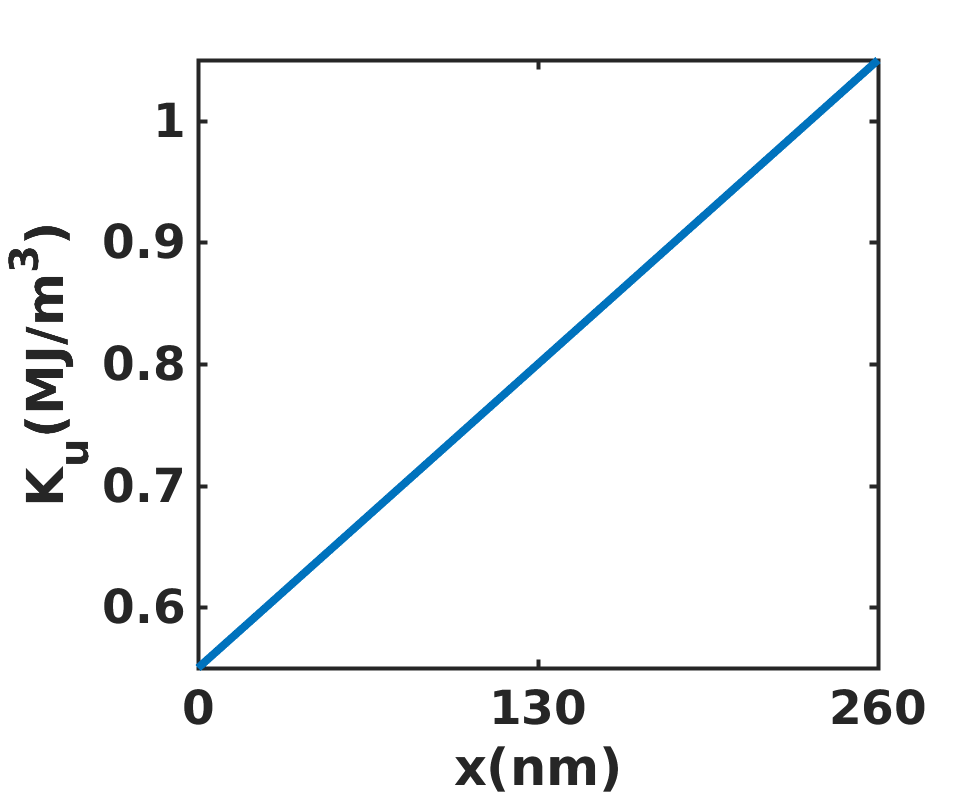}}
    \subfigure[]{\includegraphics[scale=0.12]{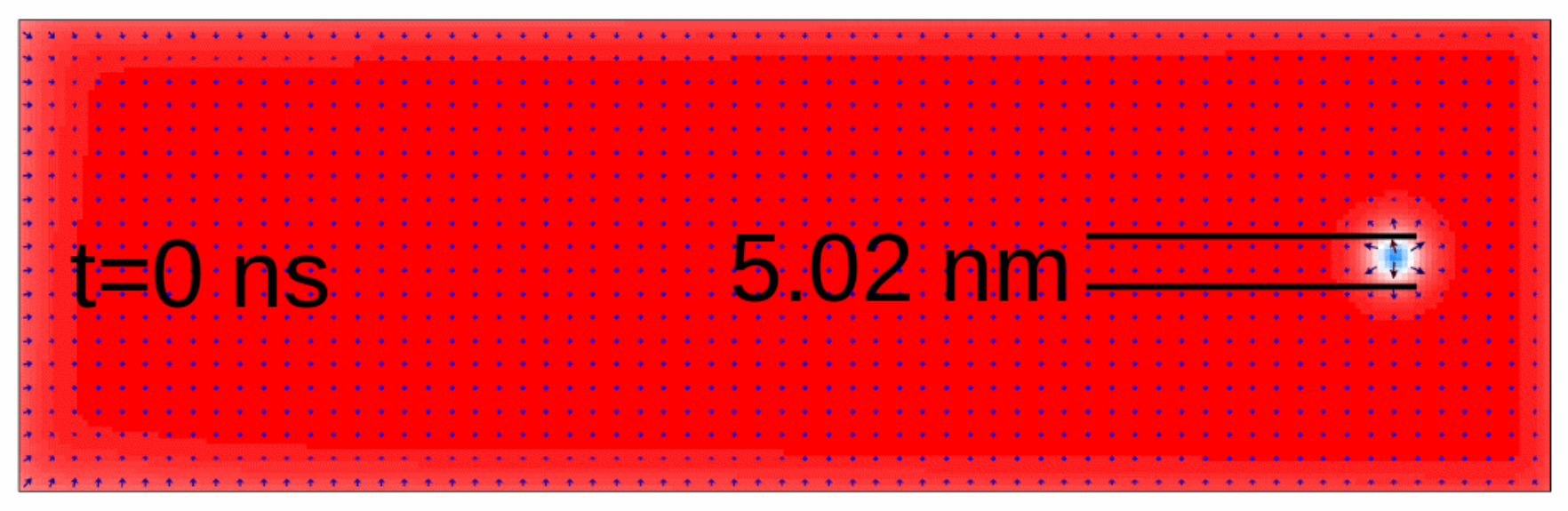}}
    \subfigure[]{\includegraphics[scale=0.12]{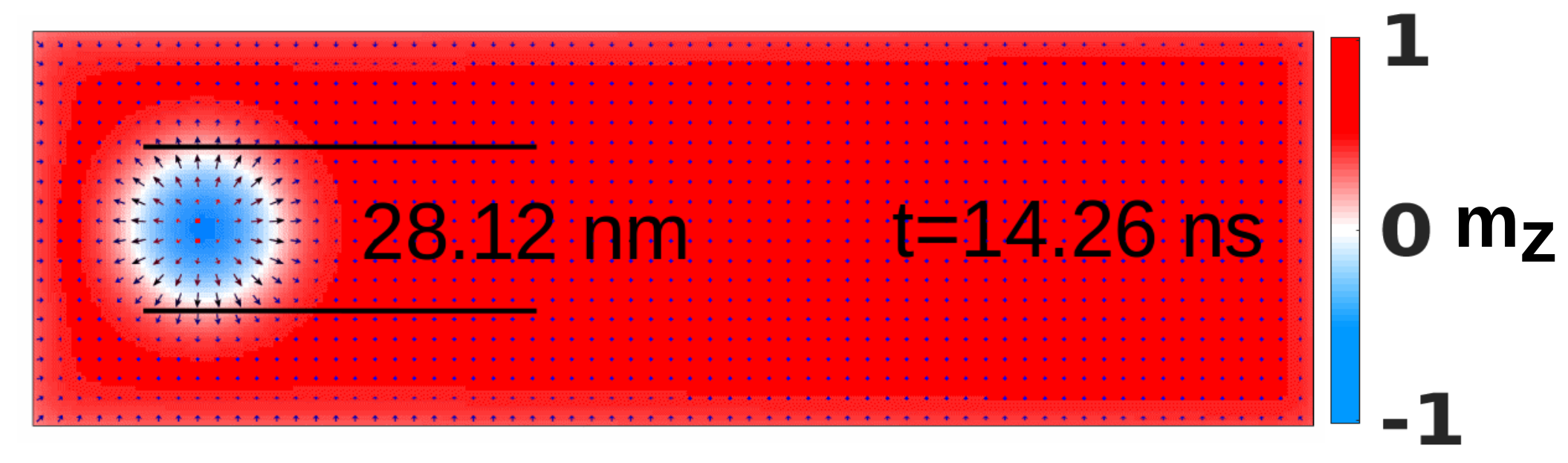}}
    \hspace*{4cm}
    \subfigure[]{\includegraphics[scale=0.25]{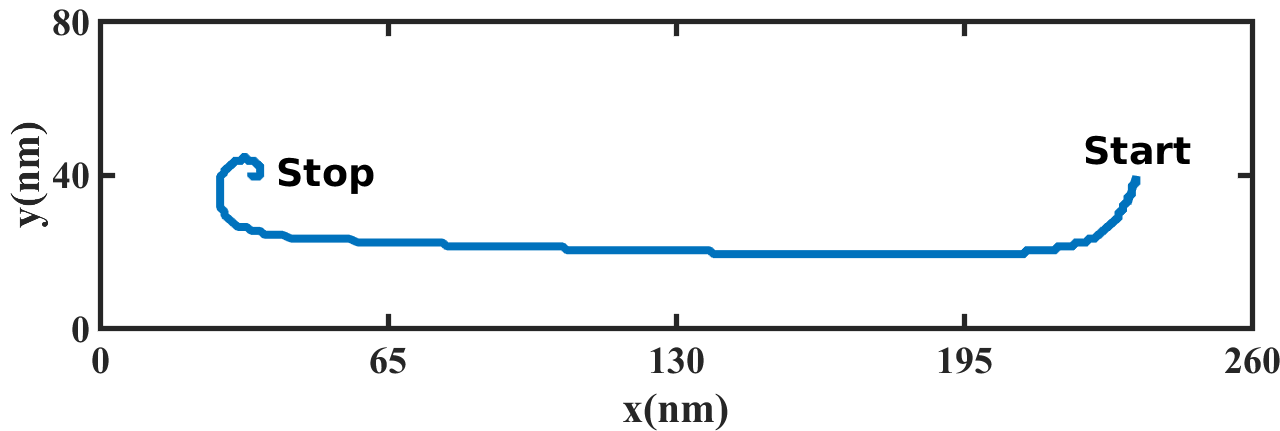}}
    \caption{Skyrmion motion under uniaxial anisotropy field gradient. (a) Schematic of the device inducing voltage controlled uniaxial anisotropy gradient along the length of the nanotrack. (b) Uniaxial anisotropy profile along the length of the nanotrack. Anisotropy constant value increasing linearly from the left to right side of the nanotrack. (c) Initial position and size of the skyrmion, (d) Final position and size of the skyrmion during its motion due to anisotropy gradient. The color bar is showing the z component of the normalized magnetization vector. (e) The path of the skyrmion motion in the x-y plane of the nanostrip .}
    \label{Kugrad}
\end{figure}

In this section, we describe the effect of anisotropy constant $K_u$ on skyrmion. It was reported that the value of $K_u$ can be changed by applying a voltage in the FM layer \cite{wang2012electric,kang2016voltage}, so we place the skyrmion in a nanostrip with an anisotropy gradient along its length, with no spin current in FM layer to investigate the sole effect of $K_u$ on skyrmion. The gradient can be obtained by varying the thickness of the insulator as shown in Fig. \ref{Kugrad}(a). The dimension of the FM layer is taken 260$\times$80$\times$0.4 $\mathrm{nm^3}$ and the cell size for the simulation is used 1$\times$1$\times$0.4 $\mathrm{nm^3}$. We assume a constant gradient of $K_u$ with a value of 1.923$\times 10^{12}$ J/$\mathrm{m^4}$ such that a linear relationship between the $K_u$ and the length of the nanostrip is maintained which in turn creates the values of $K_u$ at left and the right side of the nanostrip which are 0.55 MJ/$\mathrm{m^3}$ and 1.05 MJ/$\mathrm{m^3}$ 
respectively. The linear profile of $K_u$ is shown in Fig. \ref{Kugrad}(b). A skyrmion is initially created at the right side of the nanostrip with local injection of a spin current pulse of 0.5 ns with polarization along -z direction followed by a 1 ns relaxation to stabilize the skyrmion in the nanostrip. After the stabilization, we apply the $K_u$ gradient along the x-axis. At time t=0 ns, the diameter of the skyrmion at the right side is found to be 5.02 nm, as shown in Fig. \ref{Kugrad}(c), where the radius is measured from the center of the skyrmion to the region where $m_z$=0. Now due to the $K_u$ gradient, skyrmion starts to move towards the left region. From the second term in Eq. \ref{energy}, it is clear that total energy of the skyrmion is directly proportional to $K_u$ and as we are applying a $K_u$ gradient, so to minimize the energy, skyrmion starts to move towards the left side where $K_u$ is decreasing. The locus of the center of the moving skyrmion due to this anisotropy gradient is shown in Fig. \ref{Kugrad}(e), where the complete path traversed by the skyrmion is shown in the x-y plane. From this figure, it can be seen that at starting, skyrmion follows a curved path due to the Magnus force, which brings the skyrmion near the edge of the nanostrip. Due to the $K_u$ gradient, the skyrmion wants to move towards the left side which indicates the initial velocity at $t$=0 ns, should be along the -x direction. Thus the Magnus force $\boldsymbol{G}\times \boldsymbol{v}$ will be along -y direction, which pushes the skyrmion towards the lower side of the nanostrip as shown in Fig. \ref{Kugrad}(e). It is also noticed that during its motion, skyrmion diameter also increases as the skyrmion moves to the lower $K_u$ region. As we know that there is a repulsive force act on the skyrmion near the edges of the nanostrip, so as the skyrmion is getting bigger with time during its motion, the repulsive force pushes the center of the skyrmion away from the edge, as can be seen in Fig. \ref{Kugrad}(e). Finally, at t=14.26 ns the skyrmion stops at the left side of the nanostrip, due to balance between the edge reflection at the boundary and force due to the $K_u$ gradient. The diameter of the skyrmion at steady state becomes 28.12 nm when it stops at the left side of the nanostrip, as shown in Fig. \ref{Kugrad}(d). So, from this simulation, we can conclude that skyrmion prefers to stay in the lower $K_u$ region.

\begin{figure}[h]    
    \centering
    \subfigure[]{\includegraphics[scale=0.18]{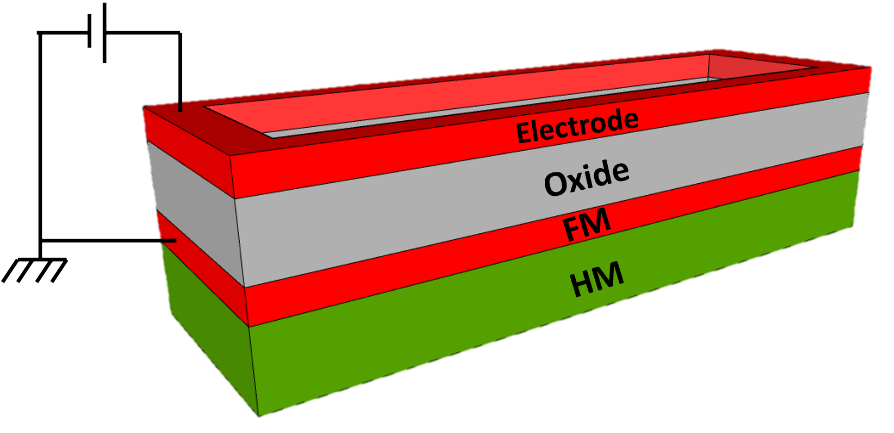}}
    \hfill
    \subfigure[]{\includegraphics[scale=0.14]{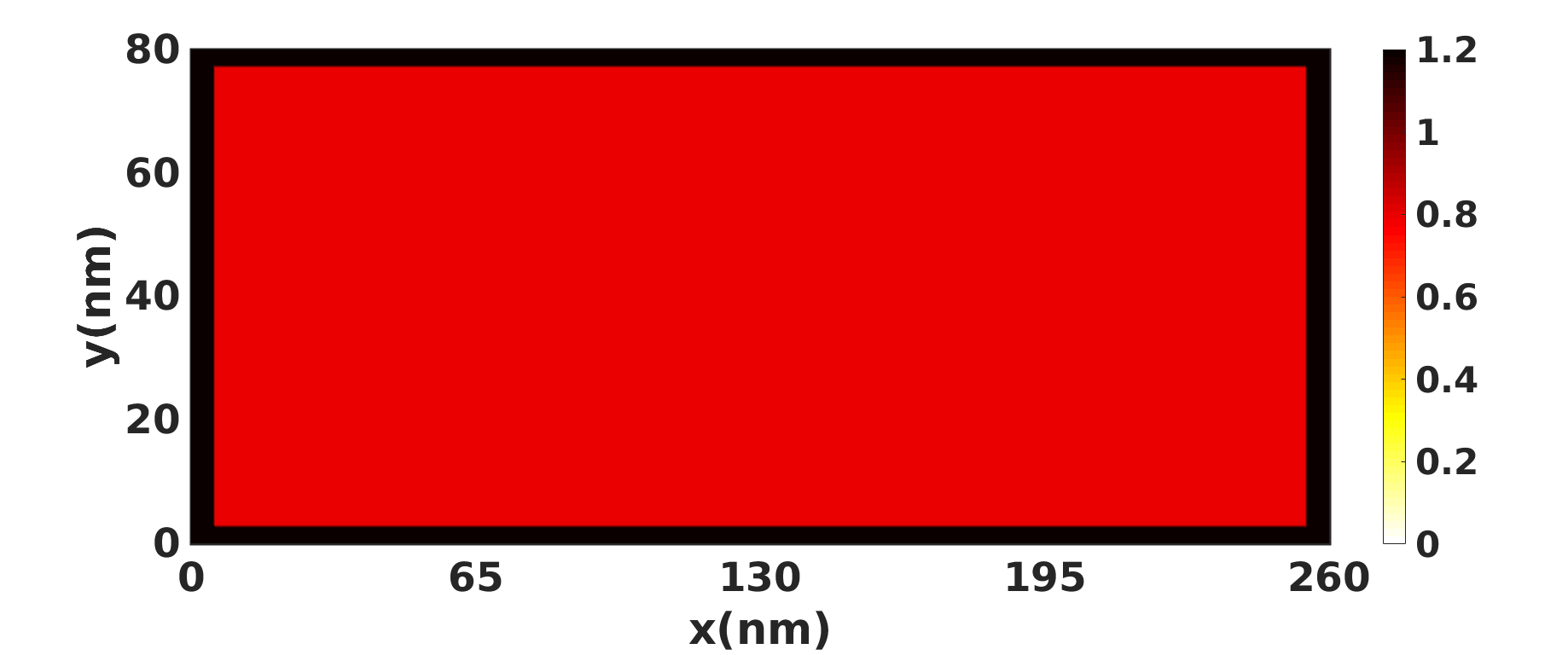}}
    \hfill
    \subfigure[]{\includegraphics[scale=0.2]{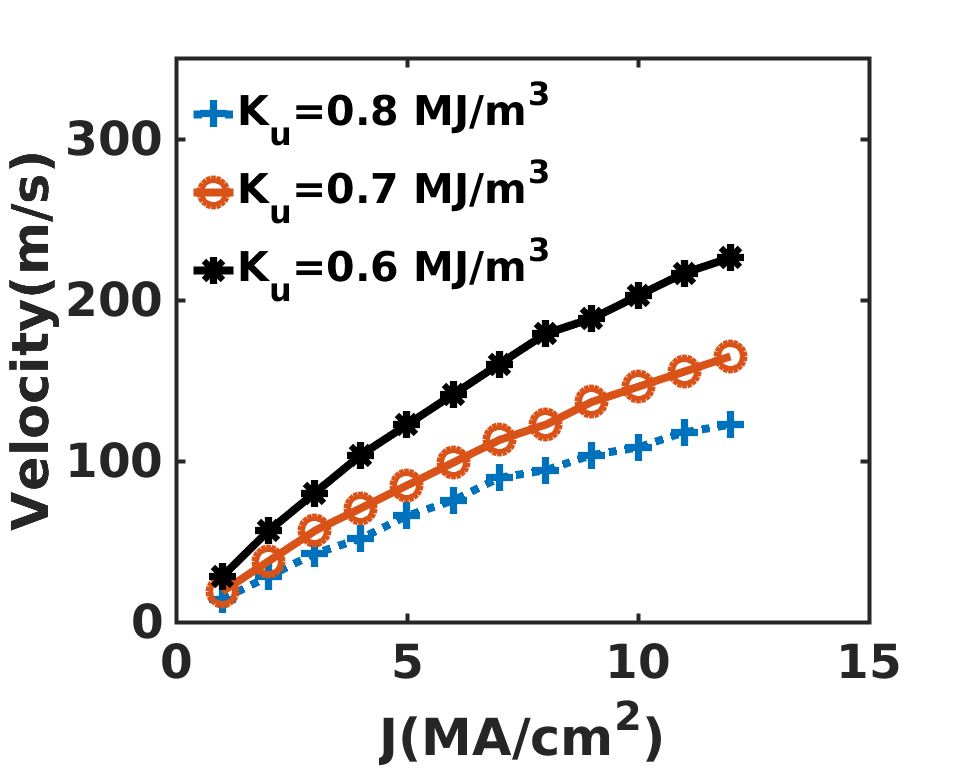}}
    
    \caption{(a) Schematic of the nanostrip with high $K_u$ barrier at the edges to stop the annihilation of the skyrmion. High $K_u$ region is created by applying a voltage to the electrode separated from the FM layer by a thick oxide layer. (b) Uniaxial anisotropy field projection of the nanostrip with high $K_u$ barrier at the edges (dark red region) and low $K_u$ (red region) at inner side. Colorbar showing the value of $K_u$ in MJ/$\mathrm{m^3}$ (c) Skyrmion velocity profile. Variation of skyrmion velocity with current density for different anisotropy constant. Velocity decreases with the increases of the anisotropy constant at any particular current density.}
    \label{Ku_vel}
\end{figure}

Next, we simulate the skyrmion motion in the nanostrip for various values of $K_u$ to investigate the effect of uniaxial anisotropy constant on the velocity of skyrmion. For driving the skyrmion in a nanostrip there are two ways to inject the spin current \cite{sampaio2013nucleation,kang2016skyrmion}, which are current-in-plane (CIP) and current-perpendicular-to-plane (CPP) configuration. In CIP configuration, a spin-polarized current is injected in-plane of the FM layer, whereas in CPP configuration, charge current is injected in the heavy metal (HM) layer beneath the FM layer which generates a spin-polarized current in the vertical direction in FM layer \cite{kang2016skyrmion}. For this simulation we take the nanostrip with the same dimension as described earlier and varied the current density ($J$) in CPP configuration, taking $K_u$ as the parameter. By the method of local spin current injection (described earlier), we first generate a skyrmion at the left side of the nanostrip and then a spin current, polarized along the -y axis is injected vertically (in z direction) in the FM layer (which can be generated by spin Hall effect) to calculate the longitudinal velocity of the skyrmion. From Ref. \cite{kang2016skyrmion} it is seen that at the higher current density ($J\geq$ 6 MA/$\mathrm{cm^2}$) skyrmion gains enough velocity to overcome the edge repulsion of the nanostrip and annihilates. As we have seen that skyrmion tends to stay at a region of lower $K_u$, so we create a region with high $K_u$ at the edges which can either be obtained by applying a local voltage as shown in Fig \ref{Ku_vel}(a) or by adding materials with high crystalline anisotropy constant, at the edges. The voltage can be applied in an electrode of width 8 nm, which is separated from the FM layer by an oxide layer with a similar width. The oxide layer is kept   
thick enough so the electron should not tunnel through. The resultant $K_u$ profile of the nanotrack is shown in Fig. \ref{Ku_vel}(b). Here, the darker red region is having a higher value of anisotropy constant which is 1.2 MJ/$\mathrm{m^3}$ \cite{sampaio2013nucleation}, and the inner region having lower values ($K_u$=0.6-0.8 MJ/$\mathrm{m^3}$), which we have taken as a parameter to plot skyrmion velocity as a function of J. Although this high $K_u$ barrier enhances the edge repulsion which helps us to increase the magnitude of $J$ to achieve higher velocity, but at a much higher current density ($J>$12 MA/$\mathrm{cm^2}$) skyrmion gets enough velocity to penetrate into the high $K_u$ region and finally annihilates at the edge. The result for velocity vs. current density is shown in Fig. \ref{Ku_vel}(c), where it is observed that skyrmion velocity increases with current density and at any fixed $J$, lower $K_u$ leads to higher velocity. 

\subsection{Oscillator}\label{osci_section}
\begin{figure}[h]
    \centering
    \subfigure[]{\includegraphics[scale=0.2]{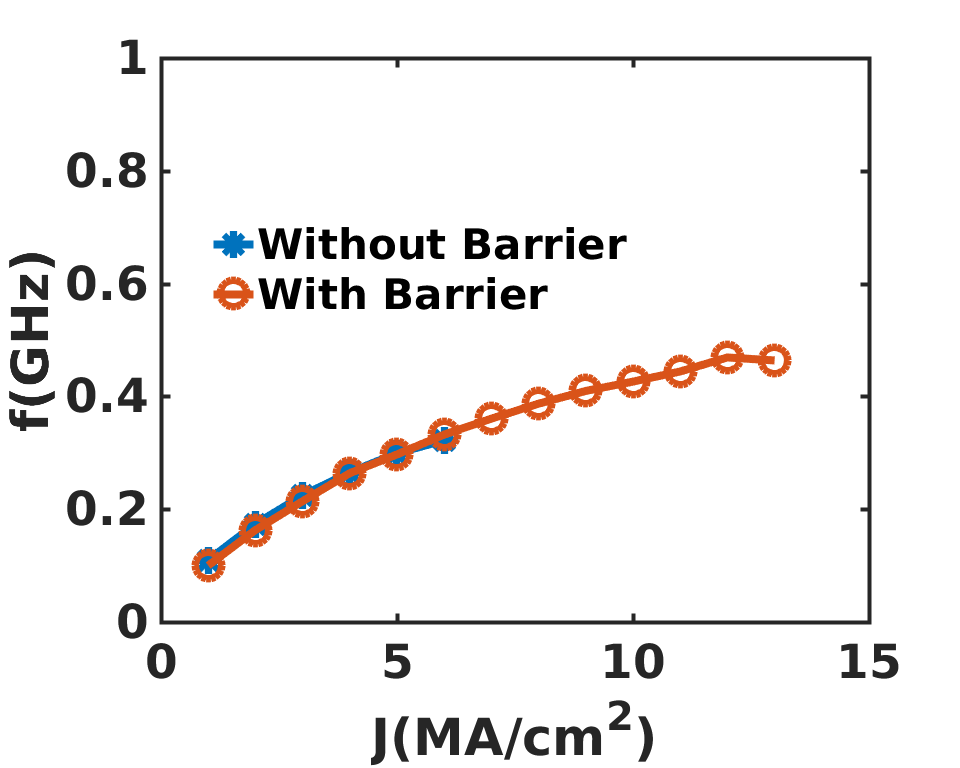}}
    \hspace*{1cm}
    \subfigure[]{\includegraphics[scale=0.2]{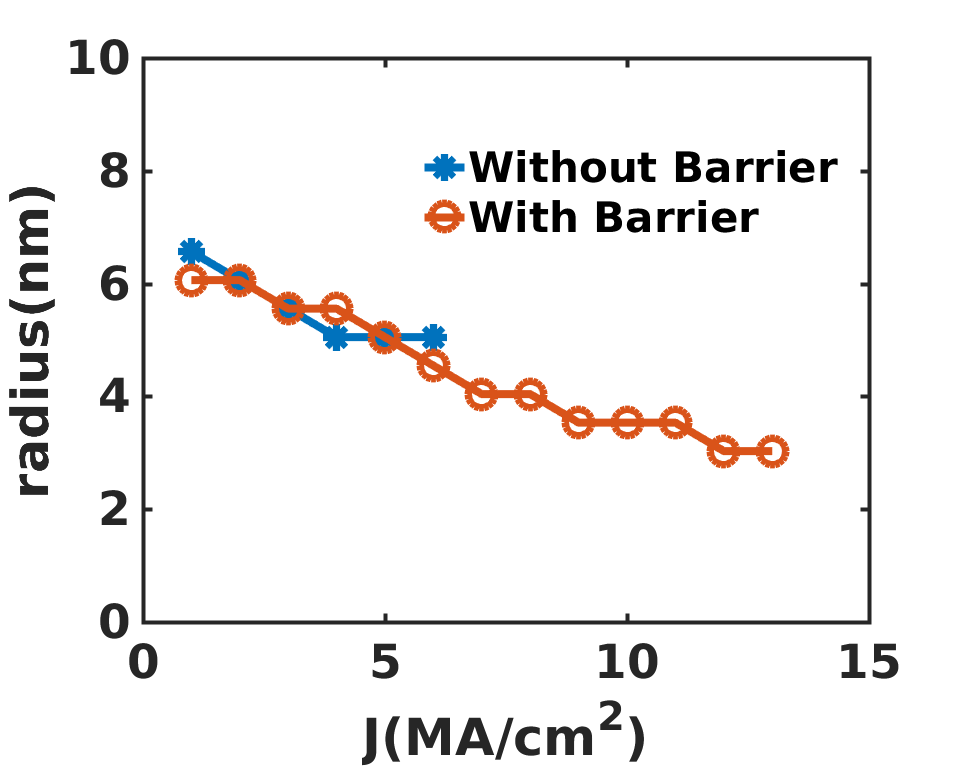}}
    
    \caption{Skyrmion gyration profile at $K_u$=0.8 MJ/$\mathrm{m^3}$ for with and without the barrier. (a) Variation of skyrmion gyration frequency with current density. (b) Variation of skyrmion radius with current density. }
    \label{freq1}
\end{figure}

In this section, we describe the detail characteristics of the skyrmion based oscillator and how to improve its performance. As the skyrmion is in the FL and the role of the RL is to provide the spin-polarized current, so the FL is only simulated in OOMMF. As discussed in Section \ref{device_structure}, we consider the device to be a circular nanodot, where the skyrmion moves in a circular path. From the result discussed in Section \ref{Ku_dendency}, it is understood that by putting a high $K_u$ barrier at the edges, 
\begin{figure}[h]
    \centering
    \subfigure[]{\includegraphics[scale=0.2]{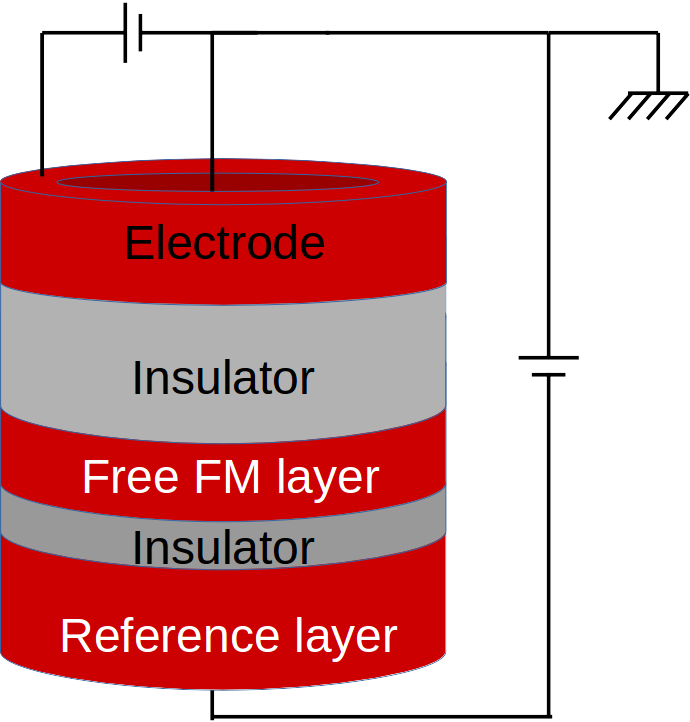}}
    \hspace*{1cm}
    \subfigure[]{\includegraphics[scale=0.25]{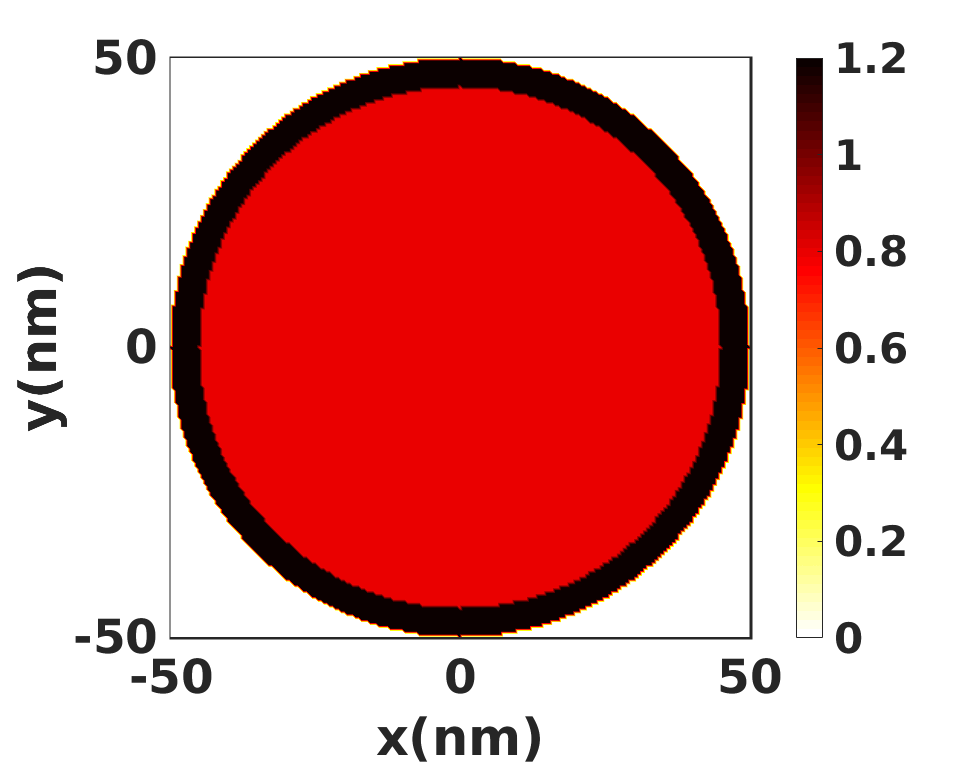}}
    
    \caption{(a) Schematic of the skyrmion based oscillator device with high $K_u$ barrier. High $K_u$ region is created by applying a voltage at an annularly shaped electrode separated by a similarly shaped thick insulator. (b)Anisotropy field projection of the circular nanodot with a high $K_u$ barrier at the edges. The darker red region represents the higher $K_u$ (1.2 MJ/$\mathrm{m^3}$) value and the red region represents the lower $K_u$ value. }
    \label{circular_barrier}
    
\end{figure}
we can increase the input current density that leads to an increase in the skyrmion velocity. As we are trying to design the skyrmion based oscillator where the skyrmion moves in a circular path in the nanodot, hence higher velocity leads to higher frequency of the skyrmion oscillator. To explore the effect of the high $K_u$ barrier on the oscillator, we consider two cases such as nanodot without high $K_u$ barrier and with high $K_u$ barrier. For the device without the high $K_u$ barrier, first, we create a skyrmion in the FL of the nanodot and applied the current to study its performance. Then we put a high $K_u$ barrier at the circumference of the nanodot of width 5 nm as shown in Fig \ref{circular_barrier} to study its effect on the frequency. High $K_u$ region is created in the same manner as discussed in Section \ref{Ku_dendency}, which can be done by applying a voltage to the annularly shaped electrode which is separated from the FL by a thick oxide layer of the same width. For this simulation, we use the anisotropy constant value $K_u$= 0.8 MJ/$\mathrm{m^3}$, but for the high $K_u$ region it is taken to be 1.2 MJ/$\mathrm{m^3}$. Other parameter values were taken the same as mentioned in Section \ref{simdetails}. For both structures i.e. with and without the high $K_u$ barrier, we change the current density and calculate the gyration frequency as shown in Fig. \ref{freq1}(a). From this plot, we observe that frequency increases with 
the current density for both structures. From Fig. \ref{Ku_vel}(c), we have seen that for any fixed value of $K_u$, the velocity of skyrmion increases with $J$. Similarly, for the circular nanodot, skyrmion velocity also increases with $J$, which in turn increases the gyration frequency. The upper limit of the applied current density is much smaller for the device without barrier than that of with barrier. For the device, without barrier, the upper limit of $J$ is found to be 6 MA/$\mathrm{cm^2}$. Upon increasing the value of $J$ beyond this limit, skyrmion gets enough velocity to overcome edge repulsion and finally annihilates at the edge. The maximum gyration frequency we obtained is 0.32 GHz at $J$=6 MA/$\mathrm{cm^2}$ for this structure. On the other hand, the device with a high $K_u$ barrier is able to protect the skyrmion from being annihilated at higher $J$ which increases the gyration frequency further. For the device with high $K_u$ barrier, we are able to increase the value of $J$ up to 13 MA/$\mathrm{cm^2}$, although the highest frequency we obtained for this structure at 12 MA/$\mathrm{cm^2}$ which is 0.47 GHz. With the increase of velocity, the Magnus force acting on the skyrmion also increases. When this force becomes larger than the edge repulsion at higher $J$, skyrmion gets annihilated. As the energy of the skyrmion becomes high at the higher $K_u$ region, so it tries to avoid this region, but with the increase of $J$, the Magnus force pushes the skyrmion towards the radial direction that forces the skyrmion to enter into the high $K_u$ region and finally annihilates at the edge. Another interesting thing we notice that with the increase of $J$, skyrmion radius becomes smaller as shown in Fig. \ref{freq1}(b). For the device without barrier, this effect is not so significant as the skyrmion annihilates at a comparatively lower value of $J$, but due to the higher range of $J$ for the device with the barrier, such effect is prominent. For the device without barrier, the skyrmion radius is 6.56 nm at $J$=1 MA/$\mathrm{cm^2}$ and for higher $J$, radius decreases due to the Magnus force and it maintains a value of 5.05 nm for $J$=4-6 MA/$\mathrm{cm^2}$. For the device with a high $K_u$ barrier, there is a significant change in radius as visible in Fig. \ref{freq1}(b). At $J$=1 MA/$\mathrm{cm^2}$, the radius is 6.06 nm, but with the increase of J, the radius decreases and the minimum radius obtained is 3.03 nm at $J$=12 MA/$\mathrm{cm^2}$. As described earlier, Magnus force increases with J which pushes the skyrmion towards the barrier, whereas due to high $K_u$ region skyrmion also feels a higher repulsive force. These two opposite forces squeeze the skyrmion, hence the radius decreases with increasing $J$. 

\subsection{Effect of $K_u$ on Frequency}\label{ku_effect}
\begin{figure}[h]
    \centering
    \subfigure[]{\includegraphics[scale=0.2]{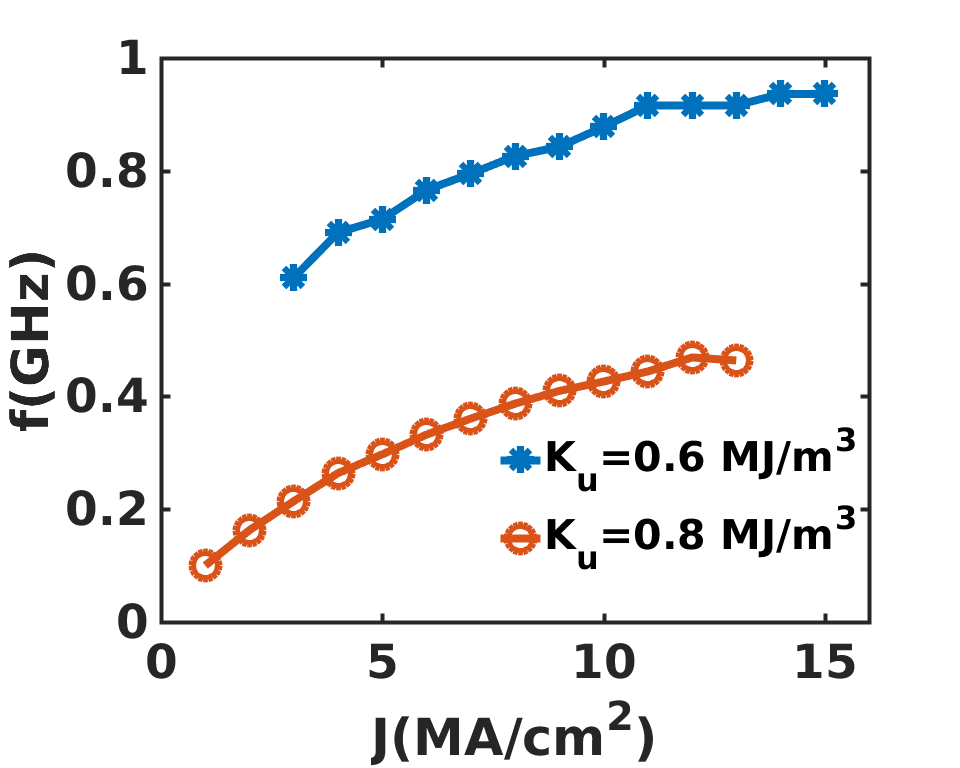}}
    \hspace*{1cm}
    \subfigure[]{\includegraphics[scale=0.2]{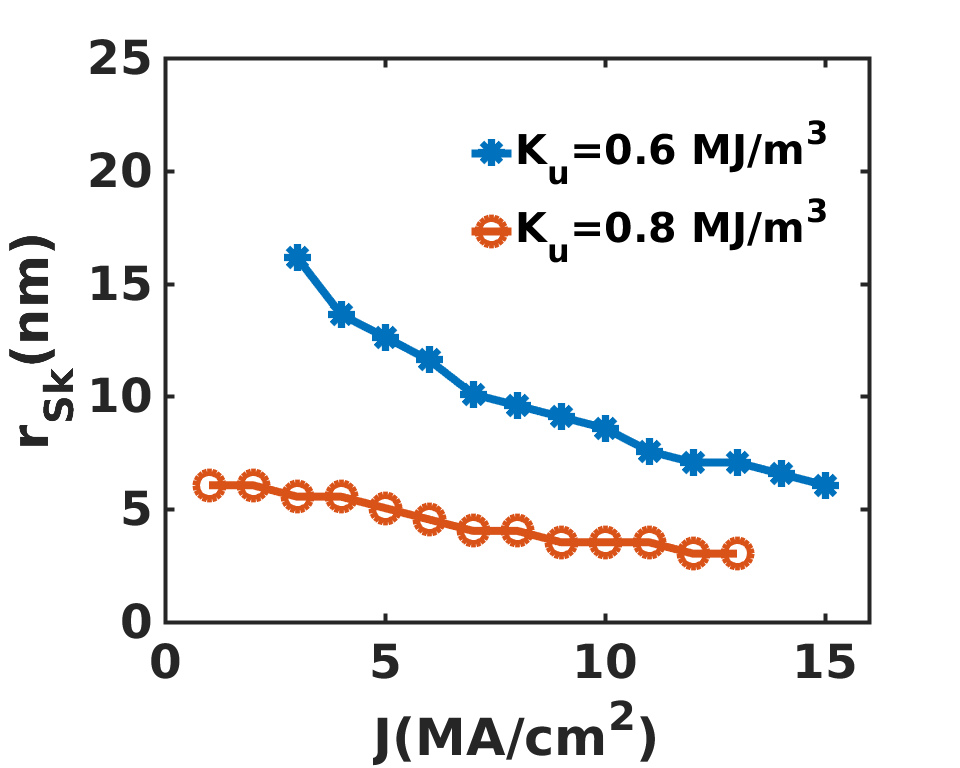}}
    
    \caption{Skyrmion gyration profile for different $K_u$ with the barrier. (a) Variation of skyrmion gyration frequency with current density. (b) Variation of skyrmion radius with current density.}
    \label{Kufrad}
\end{figure}

In this section, we describe the effect of anisotropy constant on the gyration frequency. For this simulation, we take the same device structure as described in \ref{osci_section}, but we choose two different values of $K_u$ (0.6 MJ/$\mathrm{m^3}$ and 0.8 MJ/$\mathrm{m^3}$) to investigate its effect. The simulation result for $K_u$=0.8 MJ/$\mathrm{m^3}$ has already been described in the previous section. For $K_u$=0.6 MJ/$\mathrm{m^3}$ we got similar results, but numerical values obtained for gyration frequency and radius becomes larger as shown in Fig. \ref{Kufrad}. Here, we are able to apply the current density in the range between 3-15 MA/$\mathrm{cm^2}$ 
which gives rise the gyration frequency range (0.61-0.97 GHz). The reason behind this higher values of $f$ can be explained by Fig. \ref{Ku_vel}(c), where we have seen that, for any value of $J$, a lower value of $K_u$ gives a higher velocity of the skyrmion. At lower $K_u$, as the velocity increases so it takes much lesser time to complete the circular path, which in turn increases the frequency. As shown in Fig. \ref{Kufrad}(a) that lower $K_u$ gives much higher frequency, so from the device perspective FL having lower $K_u$ is a better choice. 

On the other hand, from  Fig \ref{Kufrad}(b), we can see that lower $K_u$ leads to the skyrmion with a higher radius. The similar result we have already observed in the linear $K_u$ gradient device as shown in Fig \ref{Kugrad} (c)-(d). The highest radius we obtained is 16.16 and 6.06 nm for $J$=3 MA/$\mathrm{cm^2}$ at $K_u$=0.6 MJ/$\mathrm{m^3}$ and $J$=1 MA/$\mathrm{cm^2}$ at $K_u$=0.8 MJ/$\mathrm{m^3}$, respectively, and it starts to decrease as the J increases. For the simulation at $K_u$=0.6 MJ/$\mathrm{m^3}$, we were not able to get a steady oscillation for $J$=1-2 MA/$\mathrm{cm^2}$, where we have noticed the radius of the skyrmion becomes very large($\sim$19 nm). Due to its larger size, it feels a higher repulsive force from the edge of the nanodot, which could not be overcome at this small range of current density, which leads to the settlement of the skyrmion at the center of the nanodot within a fraction of nanosecond after starting the simulation. We also get the lowest radius of the skyrmion for these two $K_u$ values at different values of $J$. For $K_u$=0.8 MJ/$\mathrm{m^3}$, the lowest radius we get is 3.03 nm at $J$=13 MA/$\mathrm{cm^2}$, whereas for $K_u$=0.6 MJ/$\mathrm{m^3}$ it is 6.06 nm at $J$=15 MA/$\mathrm{cm^2}$. Due to the larger size of the skyrmion in lower $K_u$ device, it requires a higher Magnus force to push the skyrmion into the high $K_u$ region, which is the reason we get a slightly higher value of the upper limit of $J$ in the case of $K_u$=0.6 MJ/$\mathrm{m^3}$.

The high $K_u$ barrier discussed up to this is assumed to be very sharp (infinite gradient), but for the practical devices, it is very difficult to achieve such a sharp profile. So, we investigate the skyrmion gyration frequency in a high 
\begin{figure}[h]
    \centering
    
    \subfigure[]{\includegraphics[scale=0.25]{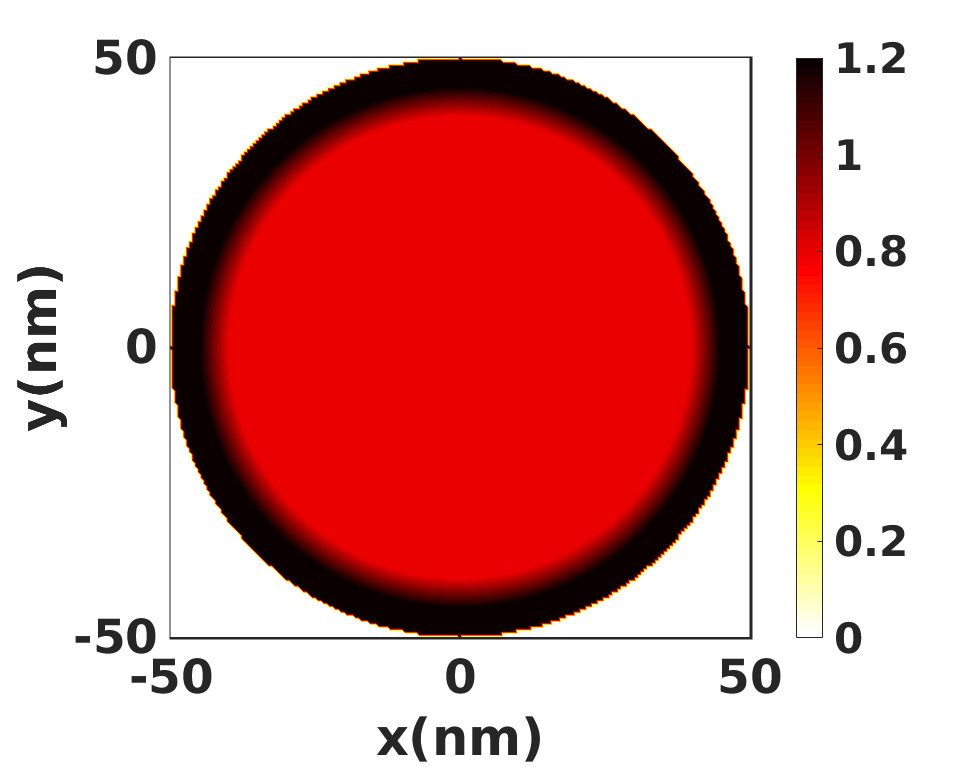}}
    \hspace*{1cm}
    \subfigure[]{\includegraphics[scale=0.25]{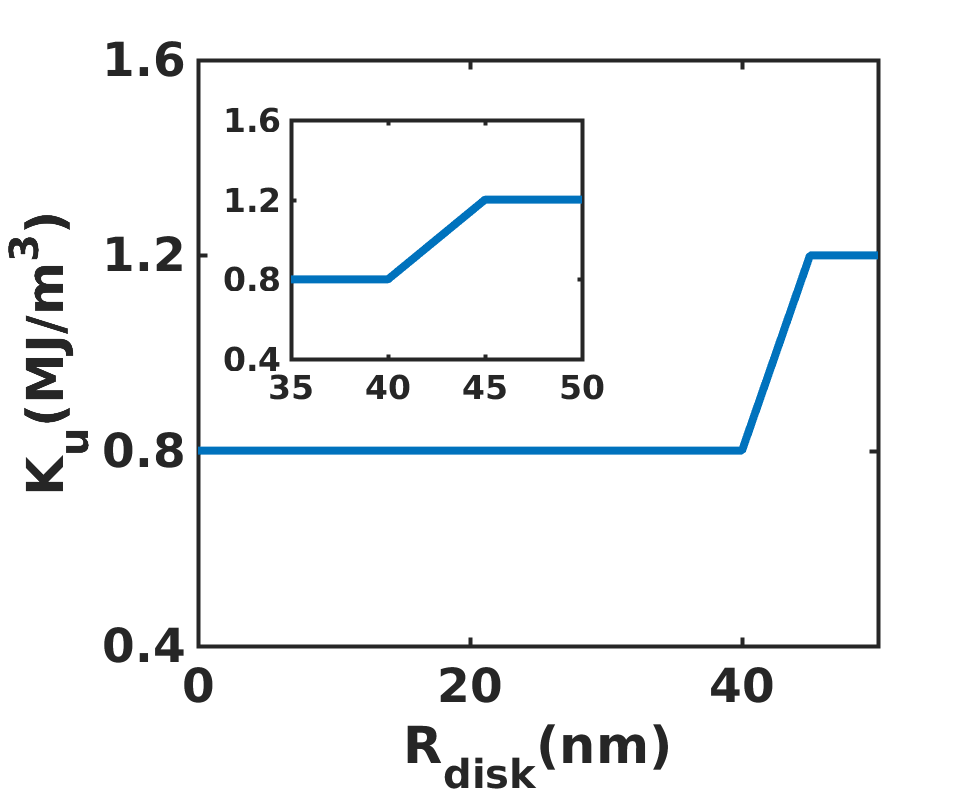}}

    \caption{ (a)Anisotropy field projection of the circular nanodot with a high $K_u$ barrier at the edges with a finite gradient. The darker red region represents the higher $K_u$ (1.2 MJ/$m^3$) value and the red region represents the lower $K_u$ (0.8 MJ/$m^3$) value and the transition from dark red to red region displays the change of $K_u$ value with a finite gradient. (b) A plot of $K_u$ profile along the nanodisk radius. The inset shows the variation of $K_u$ near the edge.}
    \label{circular_barrier_finite_grad}
    
\end{figure}
$K_u$ barrier profile with finite gradient as shown in Fig. \ref{circular_barrier_finite_grad}. As described earlier, an annular region of width 5 nm (50-45 nm) is having higher uniaxial anisotropy constant of 1.2 MJ/$\mathrm{m^3}$ inside the circumference of the nanodisk. For next 5 nm (45-40 nm) inside that annular region $K_u$ value varies from 1.2 MJ/$\mathrm{m^3}$ to 0.8 MJ/$\mathrm{m^3}$ in a linear fashion, and finally, 
\begin{table}[htbp]
    \caption{Effect of high $K_u$ barrier with infinite and finite gradient}
    \begin{center}
        \begin{tabular}{|c|c|c|c|c|}
            \hline
            \multirow{2}{*}{J(MA/$\mathrm{cm^2}$)} & \multicolumn{2}{c|}{infinite gradient} & %
            \multicolumn{2}{c|}{finite gradient} \\
            \cline{2-5}
            & f(GHz) & radius (nm) & f(GHz) & radius (nm) \\
            \cline{1-5} 
            7 & 0.361 & 4.04 & 0.391 & 5.05\\
            \hline
            8 & 0.387 & 4.04 & 0.426 & 4.54\\
            \hline
            9 & 0.410 & 3.54 & 0.451 & 4.04\\
            \hline
            10 & 0.426 & 3.54 & 0.474 & 4.04\\
            \hline
            11 & 0.444 & 3.54 & 0.496 & 4.04\\
            \hline
            
        \end{tabular}
        \label{tab1}
    \end{center}
\end{table}
the $K_u$ maintains a constant value of 0.8 MJ/$\mathrm{m^3}$, for rest of the portion inside the nanodot ($<$40 nm). In this structure, skyrmion also moves in a circular path, but the radius of the path decreases slightly, due to the preference of the skyrmion to stay at a lower $K_u$ region, which in turn increases the gyration frequency. A comparison of the gyration frequency and the skyrmion radius for high $K_u$ barrier with infinite and finite gradient is shown in Table \ref{tab1}. Here, we see that the skyrmion radius is slightly higher in the case of $K_u$ barrier with the finite gradient. For the barrier with infinite gradient, skyrmion was not able to penetrate into the high $K_u$ region near the edge in the given range of J (7-11 MA/$\mathrm{cm^2}$) whereas, for the finite gradient, skyrmion enters into the $K_u$ gradient region slightly. Due to this, the repulsive force near the boundary on the skyrmion would be more in the case of the infinite barrier which led to obtaining a smaller size of skyrmion for this case than the barrier with a finite gradient.
\begin{figure}[h]
    \centering
    \subfigure[]{\includegraphics[scale=0.2]{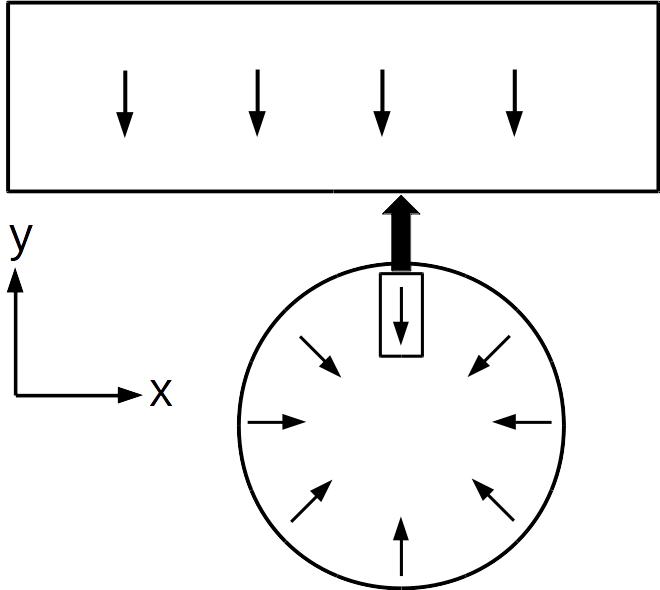}}
    \hspace*{1cm}
    \subfigure[]{\includegraphics[scale=0.2]{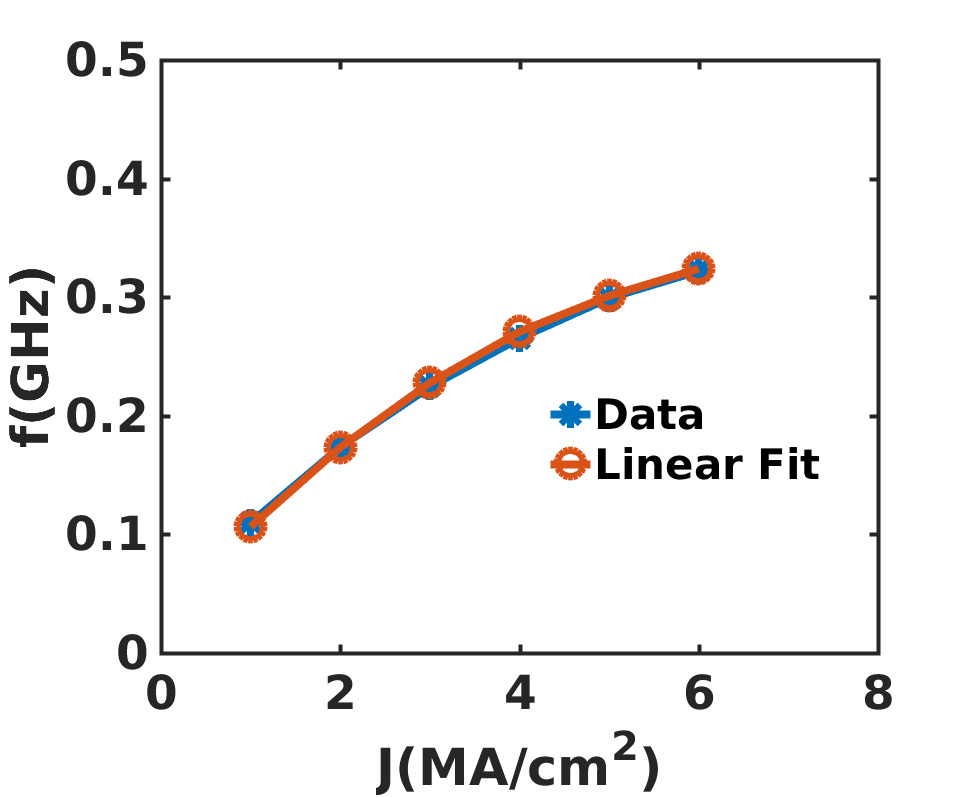}}
    \vfill 
    \subfigure[]{\includegraphics[scale=0.2]{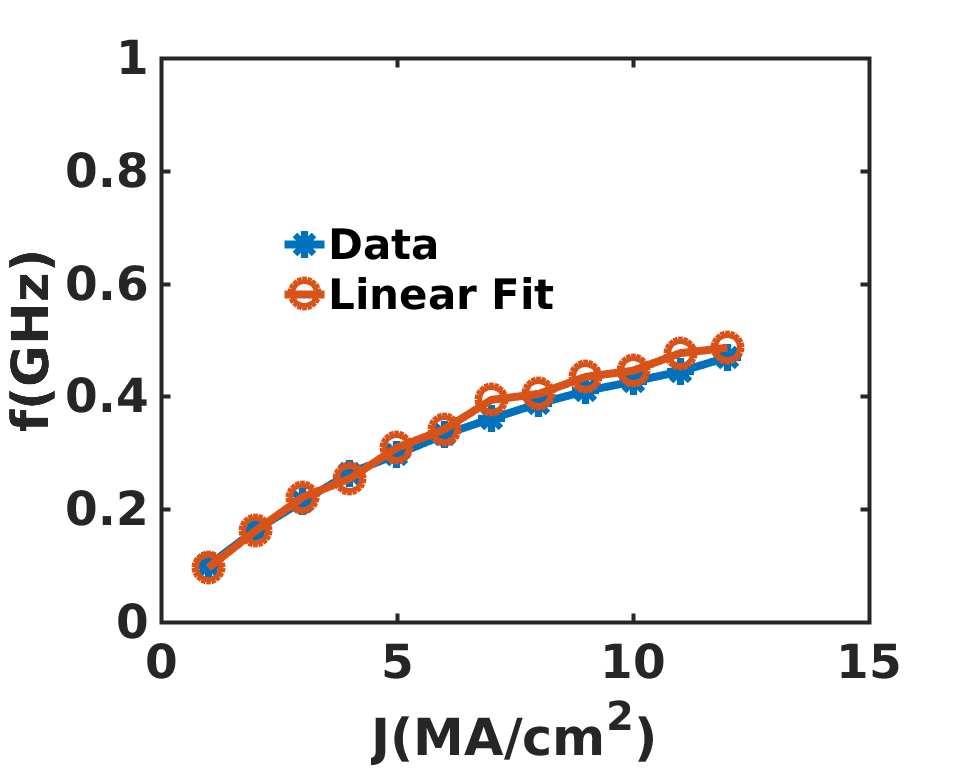}}
    \hspace*{1cm}
    \subfigure[]{\includegraphics[scale=0.21]{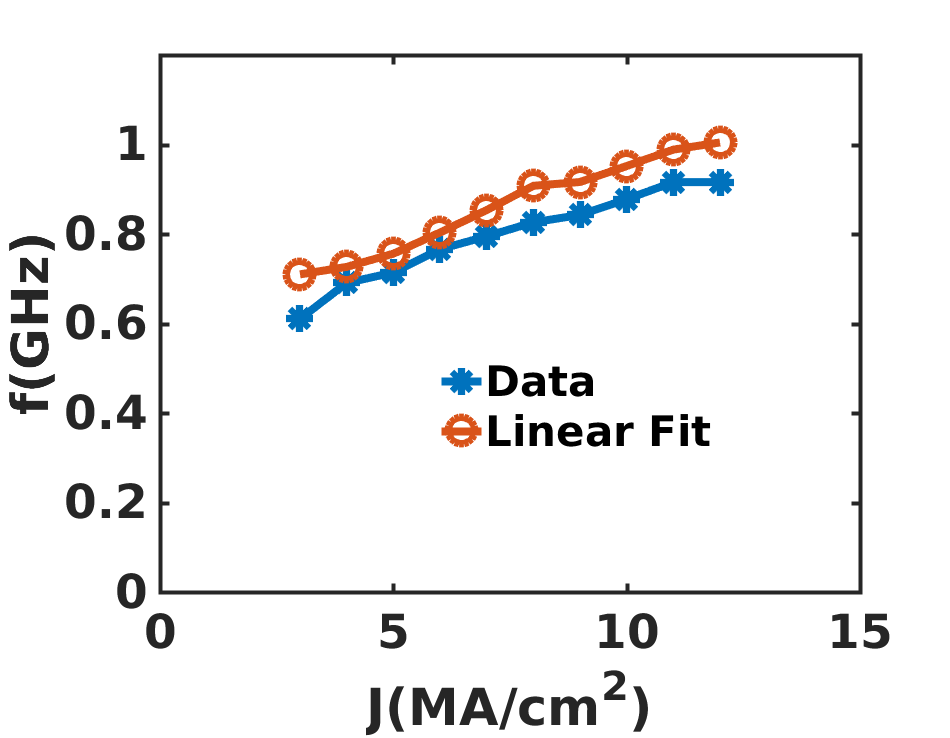}}
    
    \caption{Linear fitting of skyrmion gyration frequency. (a) Schematic view of the nanodot from the top with the spin current which takes a vortex-like spin polarization. An enlarged view of a small region of FL, where spin polarization remains uniform is shown. Variation of gyration frequency with current density for actual simulation along with the linear fitting for  without barrier at (b) $K_u$=0.8 MJ/$\mathrm{m^3}$ and  with barrier at (c) $K_u$=0.8 MJ/$\mathrm{m^3}$  and (d) $K_u$=0.6 MJ/$\mathrm{m^3}$.}
    \label{linfit}
\end{figure}

For the skyrmion motion in nanostrip, described in section \ref{Ku_dendency}, we have considered the polarization of the spin current density is along -y axis, whereas for the nanodot this direction is vortex like as shown in Fig. \ref{device}(b). Now, if we consider a very small region of the nanodot, where spin polarization can be assumed to be uniform as shown in Fig. \ref{linfit}(a), then the motion of the skyrmion in this region should mimic the motion in nanostrip and vice versa. So at steady state, one should be able to predict the frequency($f$) of the oscillator from the velocity ($v$) of the skyrmion, obtained from the nanostrip using the formula 
\begin{equation}\label{linfit_formula}
f=v/2\pi r
\end{equation}
where $r$ is the radius of the circular path in which skyrmion moves in the nanodot. We considered three cases for the simulation to verify this idea. First, we take a circular nanodot with a radius of 50 nm and a nanostrip with a dimension 260$\times$80$\times$0.4 $\mathrm{nm^3}$ with $K_u$=0.8 MJ/$\mathrm{m^3}$, but there is no such high $K_u$ barrier to protect the skyrmion. For the two structures, we calculate the velocity and the frequency of the skyrmion individually and then using Eq. \ref{linfit_formula}, we calculate the frequency from the obtained velocity. The actual frequency and the frequency obtained from Eq. \ref{linfit_formula} are plotted in Fig. \ref{linfit}(b) as `data' and `linear fit' respectively. From this figure we can see that the frequency obtained from linear fitting is in good agreement with the actual frequency, proving the validity of the idea to calculate the frequency using the linear fitting. For the second and third case, we took the nanodot and the nanostrip of the same geometrical size but with different $K_u$ which are 0.8 and 0.6 MJ/$\mathrm{m^3}$ respectively, along with high $K_u$ barrier as discussed in section \ref{Ku_dendency} and \ref{osci_section}. For $K_u$=0.8 MJ/$\mathrm{m^3}$, we have similar results which agree well with the actual frequency as shown in Fig. \ref{linfit}(c). From this figure, we can see that the frequency calculated from the linear fitting matches exactly with the actual frequency up to $J$=6 MA/$\mathrm{cm^2}$, but it deviated slightly afterward. The reason behind this small deviation can be described by the change of skyrmion size at higher current densities. As we know that at higher $J$, skyrmion gets annihilated which was prevented by keeping a high $K_u$ barrier at the edges, which in turn affects the motion of the skyrmion at the boundary between higher and lower $K_u$ region. Due to the geometrical shape difference between the nanodot and the nanostrip, the velocity of the skyrmion gets differed slightly, which leads to this deviation of the frequency obtained from the actual simulation and the linear fitting. For $K_u$=0.6 MJ/$\mathrm{m^3}$, we have done similar calculations and plotted the result as shown in Fig. \ref{linfit}(d), where we notice a larger deviation between the actual frequency and the frequency obtained from the linear fitting. From the previous results, we have seen that lower $K_u$ leads to a higher radius as well as the higher velocity of the skyrmion. Due to this larger velocity and the radius, skyrmion tends to stay near the junction between the higher and the lower $K_u$, which affects the velocity of the skyrmion at this region. In the nanostrip, the skyrmion moves in an almost straight line path after reaching near the boundary, whereas the skyrmion moves along a circular path for the nanodot. Due to the straight line motion, the velocity of the skyrmion becomes slightly higher than that of the motion in nanodot, which overestimates the frequency for the linear fitting. Due to this higher velocity, at higher $J$($>$12 MA/$\mathrm{cm^2}$) skyrmion gets into the high $K_u$ region and gets annihilated in the nanostrip and this why we have omitted the actual frequency for the nanodot for $J >$ 12 MA/$\mathrm{cm^2}$ while plotting the data in Fig. \ref{linfit}(d). 
\subsection{Effect of nanodisk radius}\label{nanodisk_diameter}
\begin{figure}[h]
    \centering
    \subfigure[]{\includegraphics[scale=0.22]{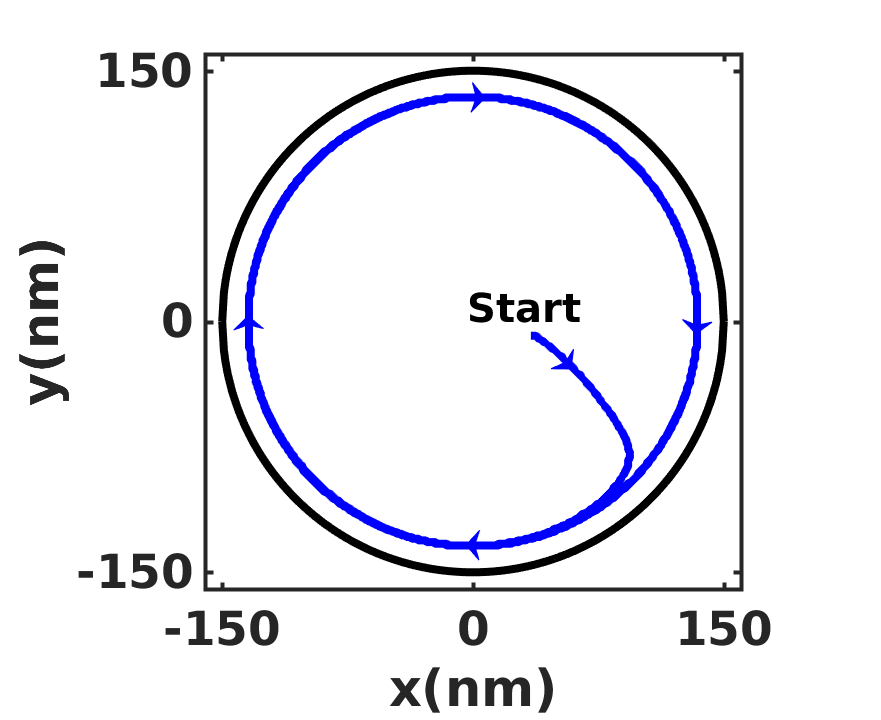}}
    \hfill
    \subfigure[]{\includegraphics[scale=0.22]{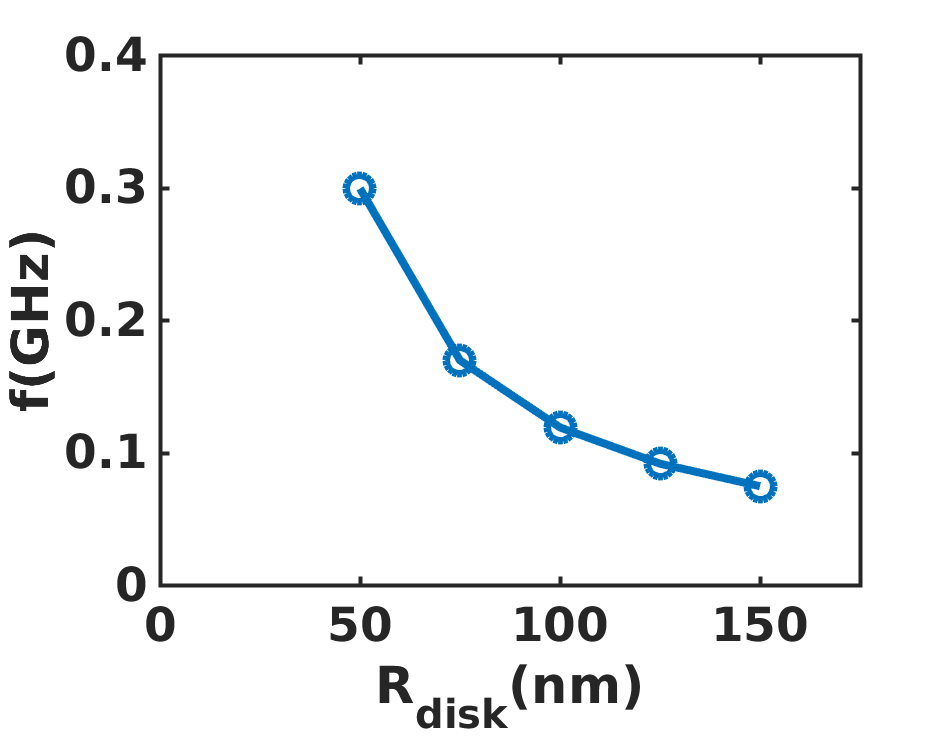}}
    \hfill
    \subfigure[]{\includegraphics[scale=0.21]{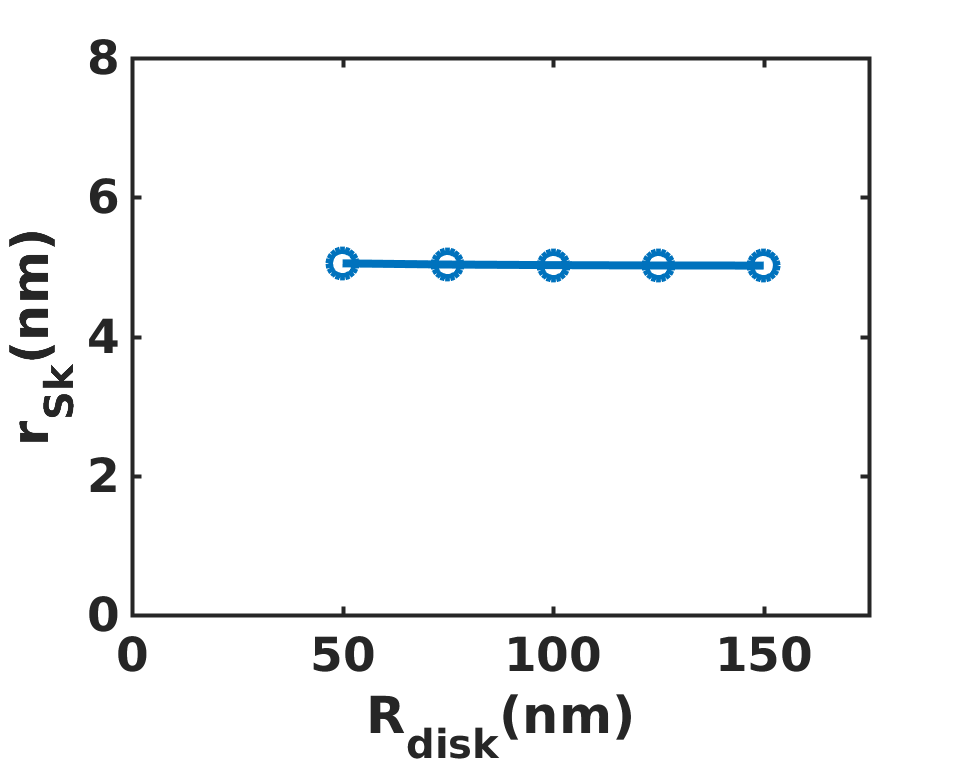}}
    
    \caption{(a) Locus of the skyrmion center(blue) of a skyrmionic oscillator in a nanodisk of radius 150 nm, whose edge is shown by the black line. (b) Variation of skyrmion gyration frequency, and (c) skyrmion radius with the radius of the nanodisk.}
    \label{disk_diameter}
\end{figure}

In skyrmion oscillator, during the motion, skyrmion starts to move along a path (shown in Fig. \ref{disk_diameter}(a) by the blue line) which is governed by the spin torque and the Magnus force acting on the skyrmion. While the skyrmion comes near the edge of the nanodisk, its Magnus force is balanced by the edge repulsion force and the skyrmion moves in 
a circular path as shown in Fig. \ref{disk_diameter}(a). From this figure, it can be understood that the gyration frequency depends on the perimeter of the circular path in which skyrmion moves, which in turn depends on the radius of the nanodisk ($R_{disk}$). So, gyration frequency should be inversely proportional to the nanodisk radius. To confirm this statement, we do the simulation by varying the radius of the nanodisk without any high $K_u$ barrier where we have kept the $J$ constant at a value of 5 MA/$\mathrm{cm^2}$. Fig. \ref{disk_diameter}(b) shows the gyration frequency decreases as we increase the radius of the nanodisk. The important result we have found that the skyrmion radius does not change with the radius of the nanodisk which is shown in Fig. \ref{disk_diameter}(c). The reason behind this phenomenon can be explained as, although nanodisk diameter increases, it does not affect the Magnus force and the edge repulsion force of the nanodisk, which keeps the skyrmion radius invariant with respect to the nanodisk diameter. From this simulation, we can conclude that the change of the nanodisk diameter affects the gyration frequency greatly but the skyrmion radius remains independent of it. 
\section{Conclusion}
We investigated the gyrotropic motion of skyrmion in a nanodot, driven by vertically injected vortex-like spin current. It is found that, in the absence of the external spin current, the skyrmion can be moved by the gradient of the uniaxial anisotropy field, and it moves towards a lower $K_u$ region. By putting a high $K_u$ barrier at the edges, we were able to protect the skyrmion from being annihilated which increase the upper limit of the injected spin current density. We have noticed that the velocity increases with the magnitude of the current density. Performing this simulation for various values of $K_u$, we have shown that the system with a lower $K_u$ leads to the higher velocity of the skyrmion. Similar to the nanostrip, by putting a high $K_u$ barrier at the edge of the circular nanodot, we were able to protect the skyrmion at higher current density, which in turn increased the upper limit of the frequency, much higher than that of obtained in the nanodot without the high $K_u$ barrier. We have shown that at a lower value of $K_u$, gyration frequency as well as the radius of the skyrmion increases. We have also shown that for the nanodot with high $K_u$ barrier, skyrmion radius reduces as the magnitude of the injected current density increases, due to balance between two opposing forces such as repulsion from the high $K_u$ region and the Magnus force. We also made a comparison on how the high $K_u$ barrier with infinite as well as finite barrier affects the gyration frequency and the radius of skyrmion. Using the velocity, obtained from the motion of the skyrmion in nanostrip, we have calculated the gyration frequency in nanodot by a linear approximation. In this method, we have noticed that this approximation works well for the skyrmion moving in nanodot with lower repulsion force from the high $K_u$ region. From this work, we can conclude that nanodot with lower $K_u$ is a better choice in terms of the frequency for the skyrmion based nano-oscillator. We have also shown that, with the increase of the nanodisk radius, gyration frequency reduces but the skyrmion radius does not change for a fixed current density.\\
\indent {\it{Acknowledgements: }} This work is an outcome of the Research and Development work undertaken in the project under the Visvesvaraya Ph.D. Scheme of Ministry of Electronics and Information Technology, Government of India, being implemented by Digital India Corporation (formerly Media Lab Asia). This work was also supported by the Science and Engineering Research Board (SERB) of the Government of India under Grant number EMR/2017/002853.

\section{References}

\bibliography{Reference}

\end{document}